# Thin-disk laser multi-pass amplifier

K. Schuhmann*[1,2], M. A. Ahmed[4], T. Graf[3], T. W. Hänsch[4], K. Kirch[1,2],
F. Kottmann[1], R. Pohl[4], D. Taqqu[1,2], A. Voß[3], B. Weichelt[3] and A. Antognini[1]
for the CREMA collaboration

1. Institute for Particle Physics, 8093 ETH Zurich, Switzerland
2. Paul Scherrer Institute, 5232 Villigen PSI, Switzerland
3. Institut für Strahlwerkzeuge, Universität Stuttgart, 70569 Stuttgart, Germany
4. Max-Planck-Institut für Quantenoptik, 85748 Garching, Germany

**ABSTRACT**

In the context of the Lamb shift measurement in muonic helium [1,2,3,4] we developed a thin-disk laser composed of a Q-switched oscillator and a multi-pass amplifier delivering pulses of 150 mJ at a pulse duration of 100 ns. Its peculiar requirements are stochastic trigger and short delay time (< 500 ns) between trigger and optical output [5].
The concept of the thin-disk laser allows for energy and power scaling with high efficiency. However the single pass gain is small (about 1.2). Hence a multi-pass scheme with precise mode matching for large beam waists (w = 2 mm) is required.
Instead of using the standard 4f design, we have developed a multi-pass amplifier with a beam propagation insensitive to thermal lens effects and misalignments. The beam propagation is equivalent to multiple roundtrips in an optically stable resonator.
To support the propagation we used an array of 2 x 8 individually adjustable plane mirrors. Astigmatism has been minimized by a compact mirror placement. Precise alignment of the kinematic array was realized using our own mirror mount design.
A small signal gain of 5 for 8 passes at a pump power of 400 W was reached. The laser was running for more than 3 months without the need of realignment. Pointing stability studies is also reported here.
**Keywords:** Thin-disk, multi-pass-amplifier, thermal lens, pointing stability

## 1. INTRODUCTION

A precise determination of proton and alpha-particle charge radii can be achieved by laser spectroscopy of muonic hydrogen (μp) and muonic helium (μHe$^+$), respectively. Muonic hydrogen is an atom formed by a proton and a negative muon. Similarly muonic helium is composed by a negative muon and a He nucleus. The muon is an elementary particle alike the electron but with 207 times larger mass and a lifetime of 2.2 μs. Because of the larger mass, the muon wave function significantly overlaps with the nucleus of the atom and therefore its energy levels depend strongly on the nuclear charge radius.

From a measurement of the 2S-2P transition in μp the CREMA collaboration has determined the proton charge radius 20 times more precisely [1,2] than from other determinations based on H spectroscopy and elastic electron-proton scattering. Yet the value obtained is, very surprisingly, seven standard deviations away from the world average. This discrepancy which was termed the "proton radius puzzle" has attracted large attention in atomic, nuclear and particle physics.

*skarsten@phys.ethz.ch; phone  +41 44 633 20 13

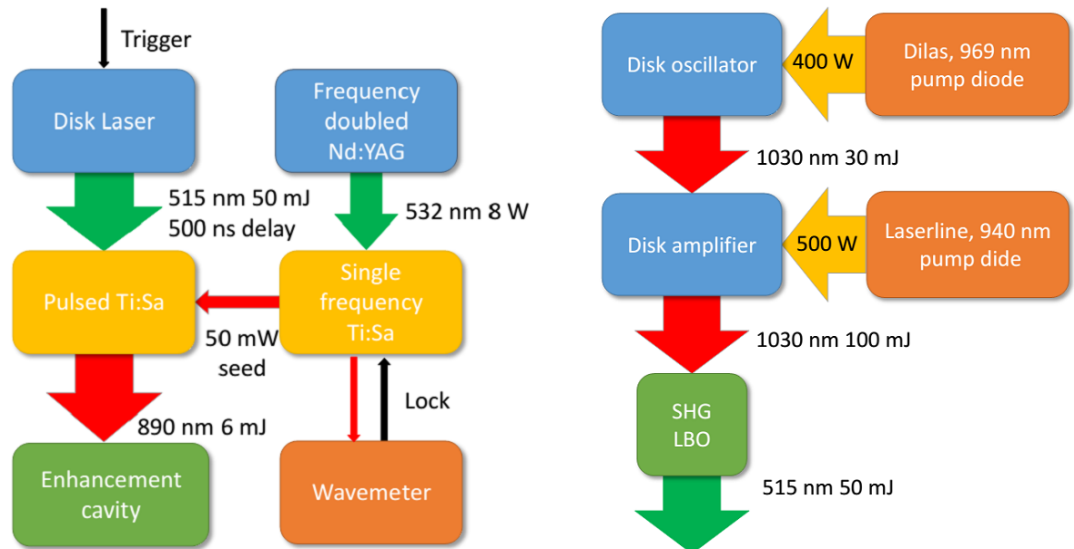

Figure 1. (Left) Scheme of the laser system. The frequency doubled thin-disk laser is used to pump a frequency locked Ti:Sa laser. The Ti:Sa pulses are successively injected into an enhancement cavity to increase fluence [7]. (Right) Thin-disk laser scheme.

Many investigations have been triggered [6], ranging from physics beyond the standard particle model, to the proton structure (low energy QCD), to atomic energy levels (bound-state QED) and several experiments have been initiated in the fields of electron-proton scattering experiment and high-precision laser spectroscopy.

To contribute to a possible solution of the "proton radius puzzle" the CREMA collaboration performed spectroscopy of μHe$^+$. The principle of the muonic He experiment is to stop muons in helium gas whereby μHe$^+$ is formed and then measure the 2S-2P energy splitting by means of pulsed laser spectroscopy. A muon entrance detector provides a trigger signal for the laser system. About 2 μs after muonic helium formation the laser pulse illuminates the muonic atom to drive the 2S-2P transition (in resonance). A scheme of the total laser system is given in Figure 1 (Left).

The continuously pumped thin-disk laser is triggered by the muon detector. Its pulses are frequency doubled (SHG) and used to pump a Ti:Sa laser, which is seeded by a stabilized continuous-wave Ti:Sa laser whose pulses are injected into a multi-pass cavity surrounding the helium gas target.

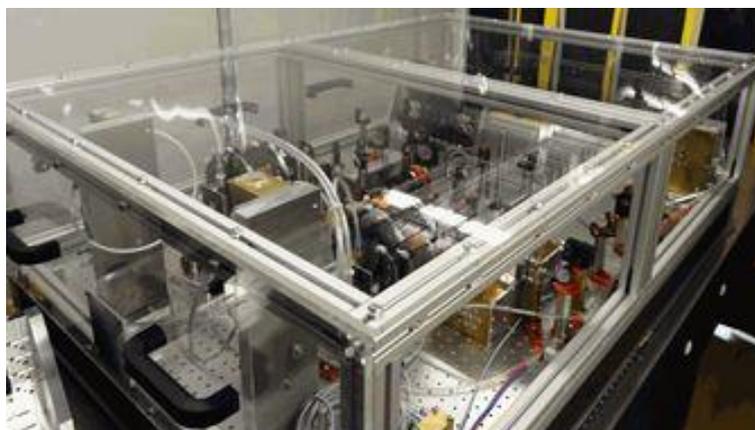

Figure 2. Thin-disk laser in the ETH laboratory developed for the *μHe$^+$* spectroscopy experiment at PSI.

As the 2S state lifetime is less than 2 μs, the laser system needs to have a short delay time between trigger and pulse delivery. Thus we have developed a thin-disk laser in an oscillator-amplifier configuration (Figure 1 (Right)) with the following requirements:

- delay between electronic trigger and laser pulse <500 ns,
- stochastic trigger,
- at least 100 mJ pulse energy,
- up to 500 Hz repetition rate.

In order to fulfill such a short delay no pulsed pumping scheme can be used. The energy needs to be stored in the laser crystal prior to trigger. We chose the thin-disk laser technology with Yb:YAG as active material. This crystal has a long upper-state lifetime of about 1 ms and can be pumped with commercially available high-power diode lasers. The thin-disk layout provides effective cooling and a small phase distortion [6]. This choice allows the extraction of pulses with large energies and high beam quality within short time from a cw pumped active material.

## 2. AMPLIFIER CONCEPT

The basic property of an optical resonator is to reproduce the identical beam shape after one roundtrip. Our amplifier design is an unwinding of the beam propagation path inside a resonator. The top panel of Figure 3 is showing a schematic of the beam routing in our multi-pass amplifier. If correct coupling is achieved, it is equivalent to the resonator shown in the bottom panel. The stability properties of the resonator are thus inherited by the amplifier. Our aim was to design an amplifier insensitive to thermal lens effects. To do this we started with the design of a resonator with the desired waist and stability.

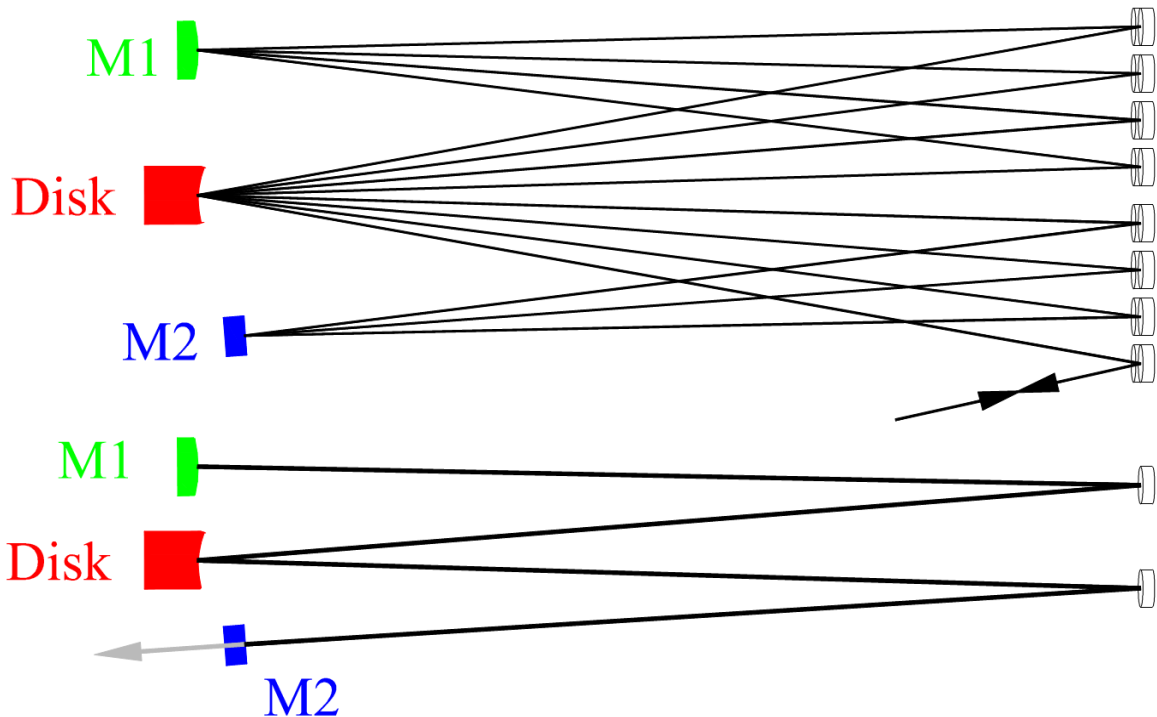

Figure 3. (Top): Scheme of the beam propagation in the amplifier. (Bottom): Scheme of the corresponding optical resonator.

In Figure 4 (Left) the beam size along the resonator is given. The resonator consists of a flat end-mirror, the thin-disk acting as a focusing mirror, a defocussing mirror and a flat end-mirror. The thin-disk position within the resonator is represented with a vertical black line. The thin-disk splits the resonator in two branches. A short-branch with a plane parallel beam and a long-branch including a Galilei telescope. Hence the long-branch is equivalent to an 11 m long free propagation. It provides thus a stable operation for a large beam size (w = 2.66 mm). The solid line represents mode beam size at the cavity design values, the dashed and dotted lines represent the beam size for variations of the thin-disks focal strength by ±0.02 diopter.

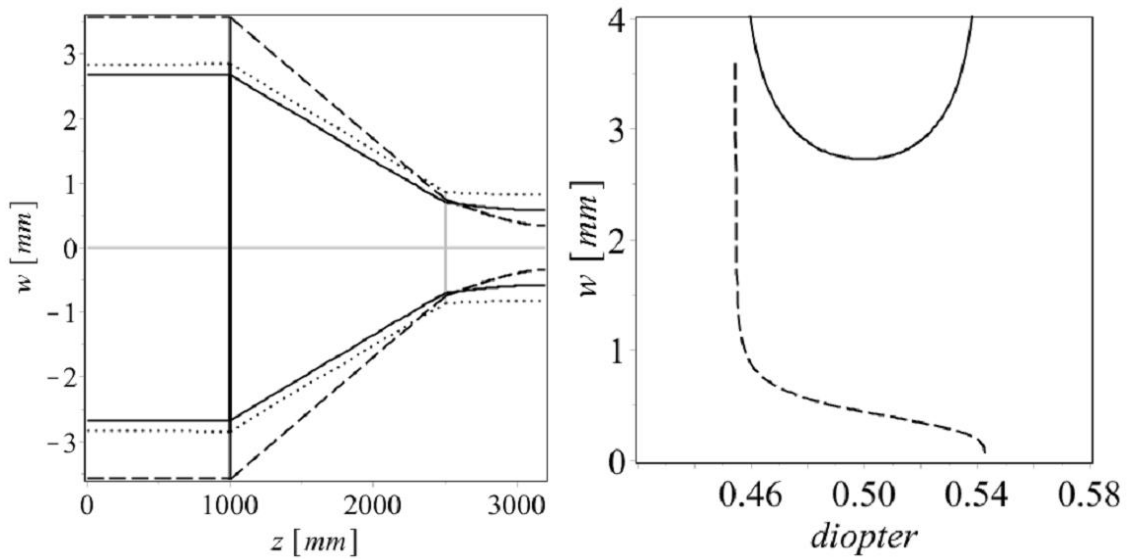

Figure 4. (Left) Beam waist along the model resonator. The thin-disk, which position is given by the black vertical lines, has a curvature of 4 m equivalent to a focal strength of 0.5 diopters. The dashed and dotted line are the beam waists for 0.52 and 0.48 diopters, respectively. (Right) Stability diagram. Beam size at the end-mirrors versus variation of the thin-disk refractive power.

Figure 4 (Right) shows the corresponding stability plot for the resonator with design value. The solid line displays the beam waist at the thin-disk and at the end-mirror of the short-branch, while the dashed line shows the mode size at the end-mirror of the long-branch. This waist is the smallest inside the cavity, consequently the most critical for mirror optical damage.

The beam size at the thin-disk position and at the short end of the cavity which is displayed in Figure 4 by the solid line does not change in first order with variation of thermal lens. A linear cavity containing a variable lens not as end-mirror always has two stability zones [8,9]. However due to the strong asymmetry of the resonator the second stability zone is at a focal strength of 2 diopters (not visible in the plot). The stability zone of a symmetric cavity has twice the width, still we chose an asymmetric layout because of our coupling scheme. The short side of the resonator is insensitive to variations of thermal lens and is suitable for beam coupling.

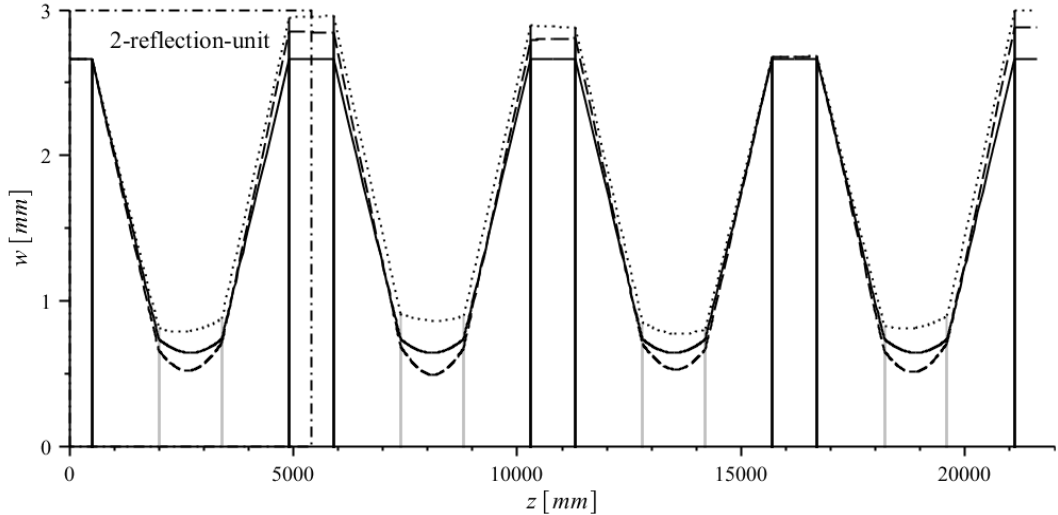

Figure 5. Beam propagation in the amplifier resulting from a concatenation of 8 optical segments equivalent to the resonator given in Figure 4.

This resonator layout is used to generate an amplifier design with 8 reflections on the thin-disk as shown in Figure 5. The laser amplifier starts with the optical system of the resonator layout of Figure 4 (Left). In order to realize a complete cavity round-trip an inverted version of this resonator is concatenated at the long-branch of the resonator. In this way a symmetric 2-reflection-unit is generated, having a short-branch on both extremities. This design provides convenient beam injection and extraction. In order to realize an 8-reflection design, 4 of these units are concatenated. The waist propagation at the design focal strength is displayed with a black line, whereas the dashed and dotted propagation have been achieved for variations of the thin-disk focal strength by ±0.02 diopter.

## 3. COMPARISON TO OTHER CONCEPTS

Our propagation layout is similar to a regenerative amplifier [10,11], as the pulse performs a limited number of roundtrips within the resonator. Within the multi-pass amplifier each roundtrips has a dedicated beam path, resulting in higher damage threshold at the cost of higher complexity of the propagation (larger number of optical elements). A multi-pass thin-disk amplifier contrary to regenerative amplifier can accept pulses of arbitrary length (from cw to fs) and can be operated also in burst mode [12].

Our design strongly differs from the 4f propagation most frequently used [13,14]. The 4f propagation is usually used to image the beam at the active medium position from one path to the next one. However the active medium (thin-disk) produces amplitude and phase distortion of the beam. Due to the imaging properties of the 4f propagation these distortions are accumulated 8 times at the same position. Hence these deviations are adding up at each pass leading to a strong optical-phase-distortion (OPD), and soft-aperture effects (see Figure 6) [15].

The collimated beam propagation used in several multi-pass designs [16,17,18,12] can be seen as a special solution for the concatenation of nearly plane-parallel resonator. Such a propagation is on the one hand very sensitive to thermal lens (see Figure 6 (Left)) and pointing instabilities, while on the other hand the propagation has the potential to be very short even for large beam waists.

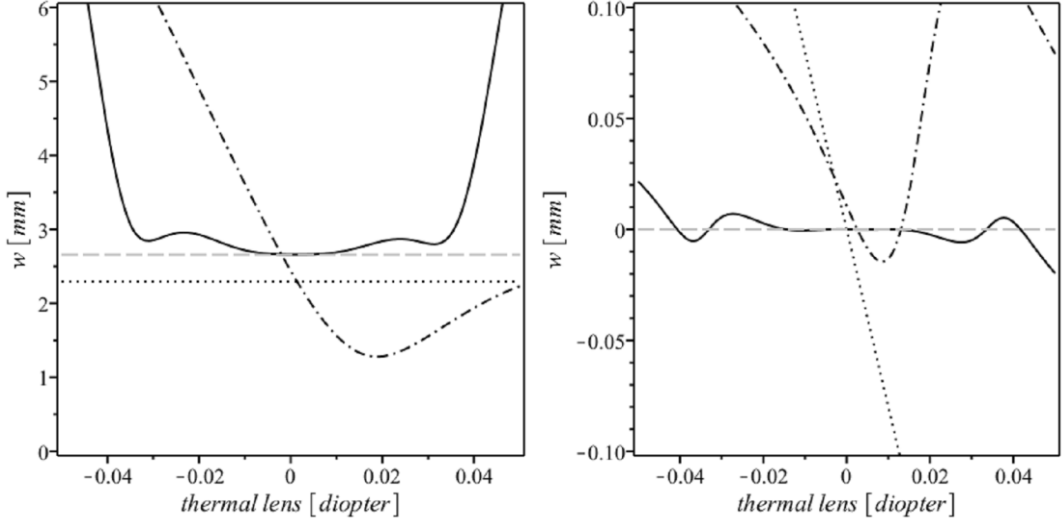

Figure 6. (Left) The exit-beam waist of the various amplifier concepts versus variation of the thin-disk focal strength. The gray dashed line represents the input beam. Our propagation (solid), 4f propagation (dotted) and the plane-parallel propagation (dashdotted). (Right) The corresponding plot for the phase front curvature.

Waist and OPD of the output beam for different propagation layouts versus variations of the thermal lens are given in Figure 6, for beam waist at the thin-disk position of 2.66 mm and 8 passes. The output beam characteristics of our multi-pass amplifier is insensitive to thermal lens variations over a range of 0.06 diopter as visible from both plots. The 4f configuration shows a stable but reduced output beam waist (due to soft aperture), while the phase front curvature (divergence) changes strongly with thermal lens. The plane-parallel design is unstable both in terms of beam waist and OPD.

The long-branch propagation between two thin-disk passes acts as a Fourier transform. All higher-order disturbances are mapped into the beam halo. This halo is cut off on the next pass due to the soft aperture of the thin-disk. The effect leads to an effective beam shaping as it takes place in a stable optical resonator. An amplifier designed for $M^2 = 1$ shows no degradation of beam quality.

The disk laser amplifier built for the measurement of muonic hydrogen [15] used soldered thin-disks which produced large higher-order diffraction artefacts. These effects reduced the beam quality already in one pass. However due to the beam forming properties the beam quality stabilized at $M^2 = 1.4$. Similar effects are expected for glued thin-disks at higher pumping power.

## 4. REALISATION OF THE MULTI-PASS AMPLIFIER

The beam routing in the amplifier is realized using an array of mirrors as shown in Figure 7. An 8-reflections amplifier requires a 16-mirror array. The multi-pass propagation was designed to allow usage of the same end-mirrors (M1 and M2) for all passes, simplifying the layout. As all array mirrors are flat and only the radius of curvature and position of M1 and M2 have to be adapted to find the correct layout, cost and alignment effort are reduced.

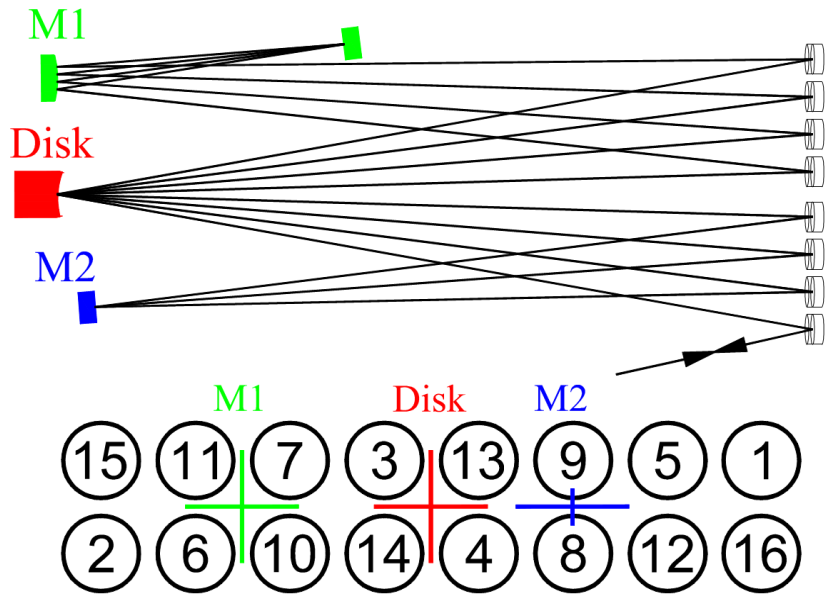

Figure 7. (Top) Scheme and propagation in the amplifier. (Bottom) A front view on the mirror array, the mirrors A1 to A16 are numbered in order of their usage. The crosses mark the points where the axis of symmetry of M1, M2 and the thin-disk hit the array plane. On the array plane these points can be seen as point reflectors, leading to the given propagation. The beam entering the amplifier over the array mirror A1 is reflected on the thin-disk to A2, from there over M1 to A3, over the thin-disk to A4, over M2 to A5 and so on.

In order to minimize the array size and hence the incident angle related astigmatism, and at the same time maximize alignment stability of the individual mirrors, we developed L-shaped mirror holder (see Figure 8). The design provides mirror-to-mirror distances similar to commercial 1" compact mount solutions but it has more precise alignment and superior stability compared to high precision mounts. This is the consequence of a longer lever (37.7 mm compared to 42.4 mm), strong springs (4 x 12 N) and precise adjustment screws (1/4"-100) mounted on a 25 mm thick aluminum plate.

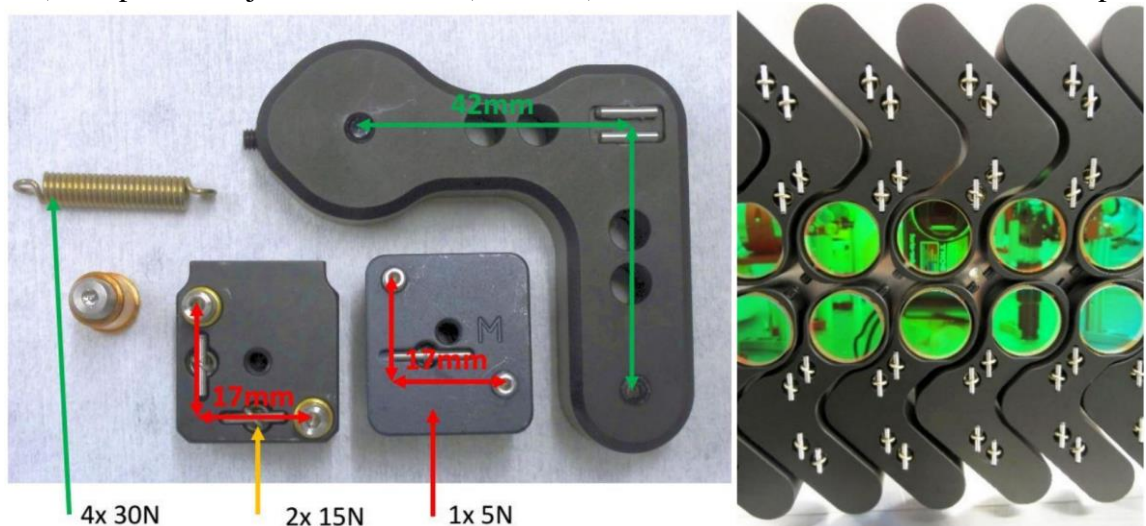

Figure 8. (Left) L-shaped mirror mount compared to commercial solutions Newport 9771 and Thorlabs KSM mounts which provide similar mirror-to-mirror distance. (Right) The mounted mirror array. The shape of the front plate was optimized for close placement. The interlaced array of L-shaped mirror mounts combine minimal placing of 31mm and maximal stability.

## 5. MEASUREMENT OF LASER OPERATION

The beam quality of the system is excellent: the $M^2$ value was measured to be 1.00(3) for all passes using a Thorlabs Beam Profiler. An output energy of 145 mJ was observed. To avoid optical damage during the 2 years of data taking, the system was operated at an output energy of about 90 mJ.

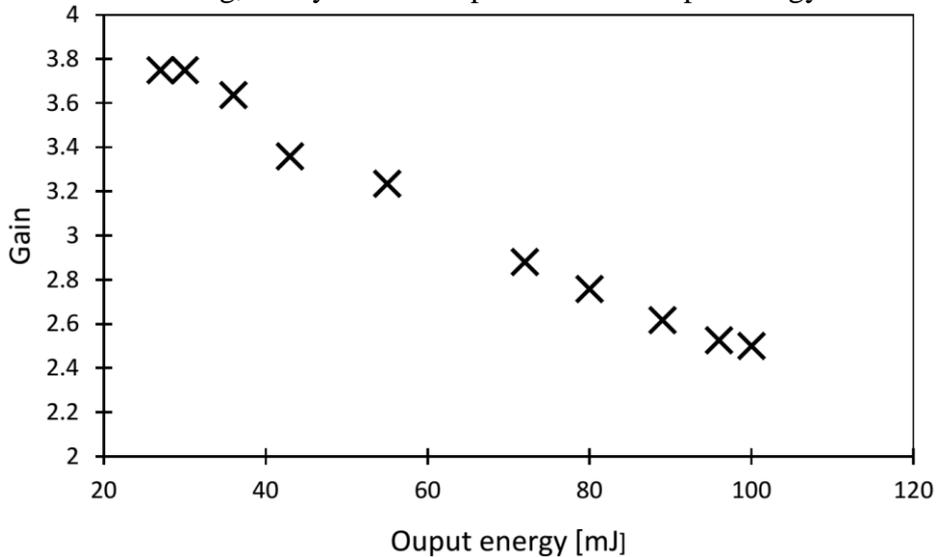

Figure 9. Amplifier gain versus output energy. The small signal gain is 5.1.

Figure 9 shows the amplifier gain as a function of the output energy. Figure 10 shows output energy as a function of input energy taken 145 days apart without realignment of the amplifier. The gain of the amplifier remained basically unchanged over weeks of continuous operation as well as 3 months of shutdown including the air condition with room temperatures reaching temperatures well above 30°C and several cooling water shutdowns. The minimal difference of the two plots shows the long-term stability of the amplifier.

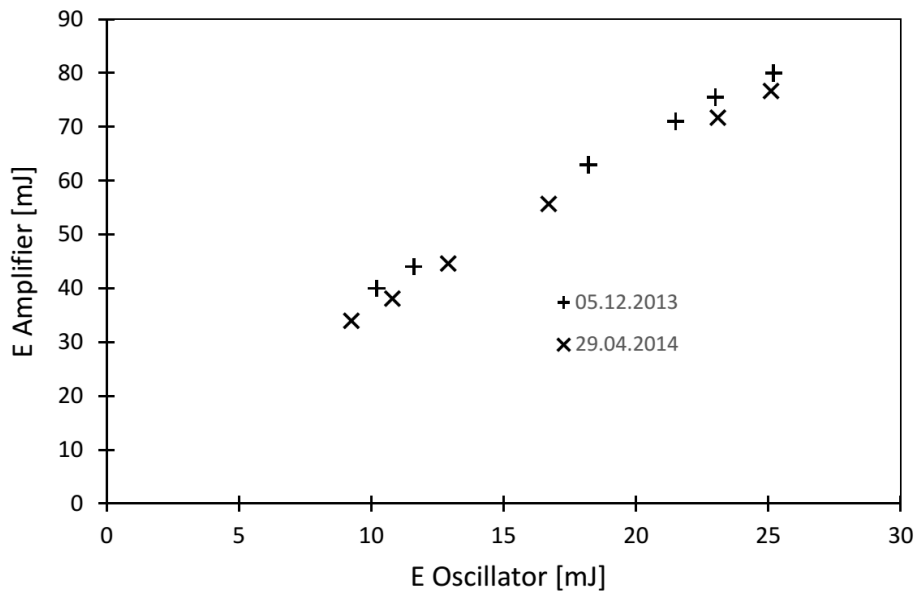

Figure 10. Output energy of the amplifier as function of input energy.

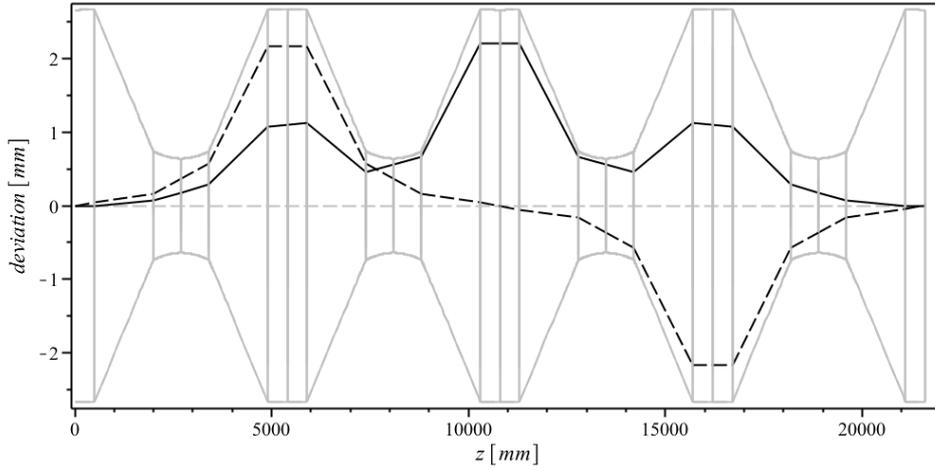

Figure 11. Misalignment plot. The deviation of the beam from the optical axis for a tilt of the in-coupled beam by 0.0025 mrad (solid) and for a tilt of the thin-disk by 0.01 mrad (dashed) is given. The Gaussian beam waist evolution along the amplifier path is shown for comparison (gray).

## 6. MISALIGNMENT SENSITIVITY

In this chapter we present a theoretical study of misalignment and pointing instabilities for various multi-pass amplifiers. These results are then compared to measurements. Figure 11 is a misalignment plot, where the deviation of the beam from the optical axis for a tilt of the in-coupled beam (dashed) and for a tilt of the thin-disk (solid) is shown.

From the dashed line behavior it can be inferred that the 2-reflection-unit acts as Fourier transform optic. After 4 of these units the output beam has identical position and pointing as the input beam. A misalignment of the thin-disk or of the end-mirrors compensates itself after 8 passes. Thus excursion from the design position will be zero even if the input beam or the thin-disk are tilted due to pointing fluctuations. In addition the output beam angle after 8 passes is not affected by tilts of the thin-disk.

As a comparison we performed a simulation study for the multi-pass amplifier of [18] where an end-mirror was replaced by a pair of 45° mirrors. This retro-reflection is causing an inversion of angle and excursion in the yz-plane while the propagation in the xz-pane stays unaffected. This vertical retro-reflector significantly increases the pointing stability in vertical direction, especially for the nearly plane parallel setup of Figure 12 (middle). The use of a corner cube reflector with 3 mirrors used at an angle of 54 ° (Figure 12 (Right)) would provide the same increase of stability for both directions (vertical and horizontal).

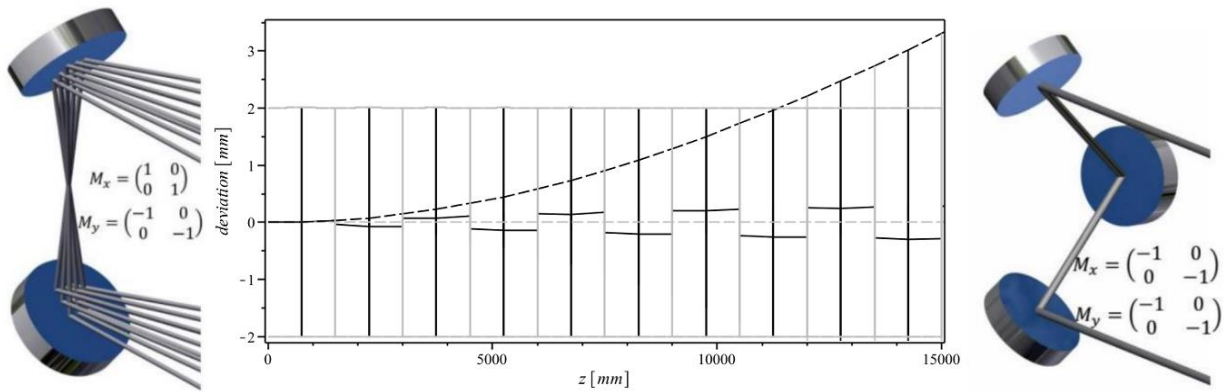

Figure 12. (Left) A pair of 45 ° mirrors used as M2 retro reflector. $M_x$ and $M_y$ are the optical matrices describing the retro-reflection. (Middle) Misalignment plot of the plane-parallel propagation used in [18]. The thin-disk positions are represented by black vertical lines. The thin-disk has a misalignment of 0.025 m rad. The dashed line shows beam excursion at a plane end-mirror while the solid line represents the beam excursion if a pair of 45 ° mirrors is used. (Right) Three mirror corner cube reflector. Using this reflector as M2 would provide stabilization in both directions.

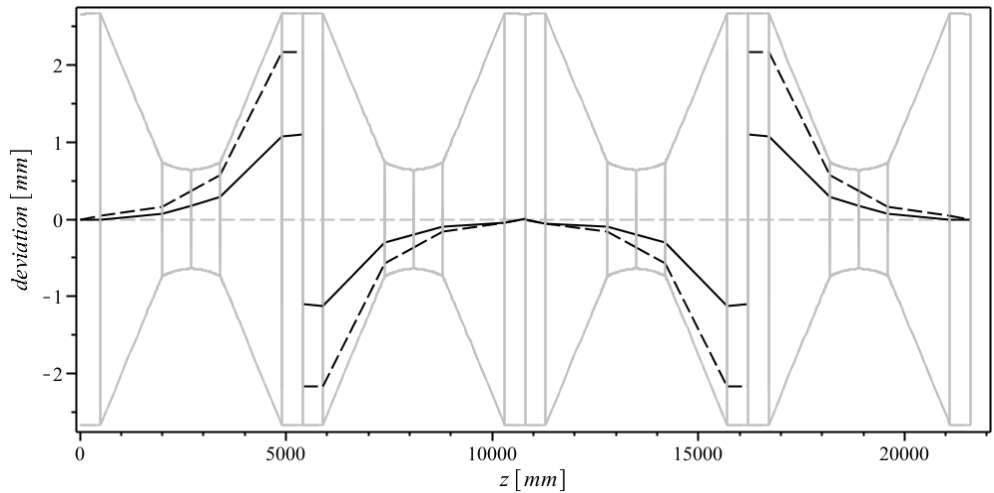

Figure 13. Misalignment plot of the modified version of our amplifier similar to Figure 11. Here M2 is replaced by a pair of mirrors as in Figure 12 (Left).

In order to determine the misalignment stability we conducted measurements at small signal operation. We used well defined forces to realize controlled pitch misalignments of the thin-disk. Tilts of 5.7 µ rad / N have been measured.

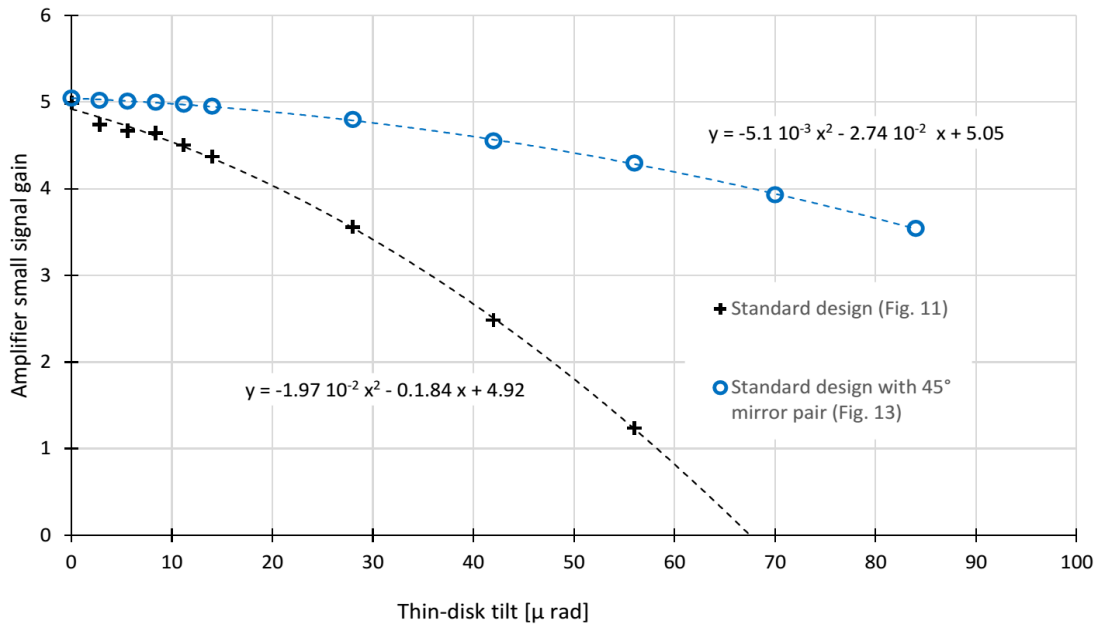

Figure 14. Amplifier gain versus thin-disk tilt.

As visible from Figure *14*, the decrease of gain versus tilting shows a quadratic dependency. The value of the quadratic term is reproducible for different alignments. The quadratic term of the propagation using a 45 ° mirror pair is 4 times smaller than the value of our standard design corresponding to a factor of two smaller misalignment sensitivity as expected (see Figure 13).

## 7. SUMMARY

We have developed a multi-pass amplifier with small sensitivity to thermal lens and pointing fluctuations. This amplifier has been used for months without the need for realignment. The introduction of a retro reflector in the multi-pass-amplifier reduced the effect of thin-disk tilt by a factor of 4.

This work is supported by the SNF_200021L-138175, DFG_GR_3172/9-1 and the
ERC StG. #279765.